\documentclass[transmag,letterpaper]{IEEEtran}
\usepackage{nopageno}
\newif\ifpeerreview

\peerreviewfalse

\newcommand{\paperID}{71}

\usepackage{lineno}

\usepackage{ctable}
\usepackage{float}

\usepackage{amsmath, amssymb}
\usepackage{dsfont}

\usepackage{cite}
\usepackage{url}
\usepackage{lipsum}
\usepackage{graphicx} 
\graphicspath{{figures/}} 

\ifCLASSINFOpdf
\else
\fi

\begin{document}

\ifpeerreview
  \linenumbers
  \linenumbersep 5pt\relax
\fi 

\title{Data-Driven Design for Fourier Ptychographic Microscopy}


\ifpeerreview
\author{Anonymous ICCP 2019 submission \\
Paper ID \paperID}
\else

\author{Michael Kellman$^{\star}$, Emrah Bostan, Michael Chen, Laura Waller\\
University of California, Berkeley\\
Berkeley, CA, USA\\
{$^{\star}$\tt\small kellman@berkeley.edu}


}
\fi

\ifpeerreview
\markboth{Anonymous ICCP 2019 submission ID \paperID}%
{}
\else
\fi

\newcommand{\lbr}{\left\lbrace}
\newcommand{\rbr}{\right\rbrace}
\newcommand{\xibf}{\boldsymbol{\xi}}
\newcommand{\Fo}{\mathcal{F}}
\newcommand{\cj}{\mathrm{i}}
\newcommand{\R}{{\mathcal R}} 
\newcommand{\x}{{\mathbf x}} 
\newcommand{\rcoord}{{\mathbf r}} 
\newcommand{\y}{{\mathbf y}} 
\newcommand{\A}{{\mathbf A}} 
\newcommand{\C}{{\mathbf C}}

\newcommand\mc{\multicolumn}

\IEEEtitleabstractindextext{%
\begin{abstract}
    Fourier Ptychographic Microscopy (FPM) is a computational imaging method that is able to super-resolve features beyond the diffraction-limit set by the objective lens of a traditional microscope. This is accomplished by using synthetic aperture and phase retrieval algorithms to combine many measurements captured by an LED array microscope with programmable source patterns. FPM provides simultaneous large field-of-view and high resolution imaging, but at the cost of reduced temporal resolution, thereby limiting live cell applications. In this work, we learn LED source pattern designs that compress the many required measurements into only a few, with negligible loss in reconstruction quality or resolution. This is accomplished by recasting the super-resolution reconstruction as a Physics-based Neural Network and learning the experimental design to optimize the network's overall performance. Specifically, we learn LED patterns for different applications ({\it e.g.} amplitude contrast and quantitative phase imaging) and show that the designs we learn through simulation generalize well in the experimental setting. Further, we discuss a context-specific loss function, practical memory limitations, and interpretability of our learned designs.
\end{abstract}

\ifpeerreview
\else
\fi
}

\makeatletter
\g@addto@macro\@maketitle{
    \begin{figure}[H]
    \setlength{\linewidth}{\textwidth}
    \setlength{\hsize}{\textwidth}
    \centering
    \includegraphics[width=18.19cm]{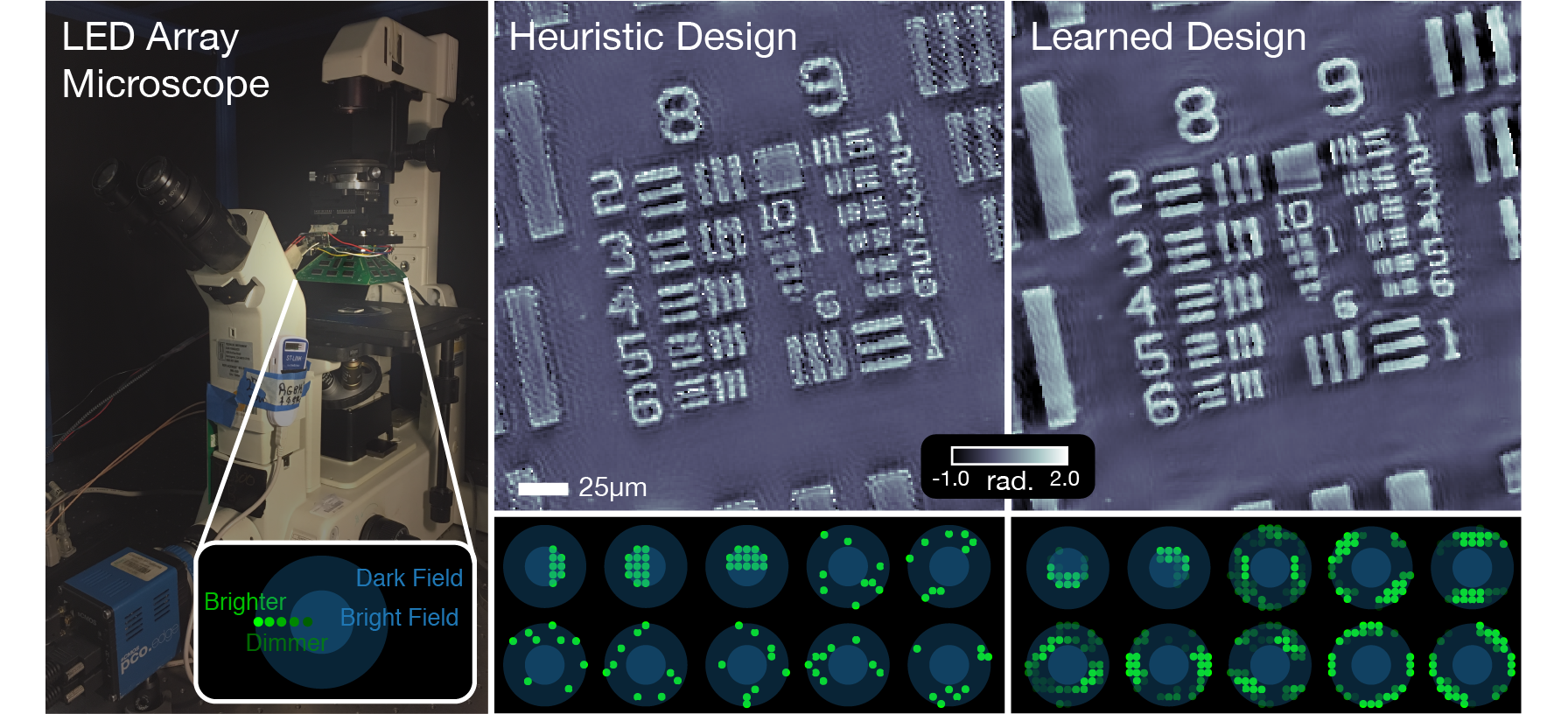}
    \caption{Fourier Ptychographic Microscopy (FPM) achieves super-resolution Quantitative Phase Imaging (QPI) with an LED array microscope (left). Here, we show the super-resolved QPI reconstruction for heuristic multiplexing designs (middle) and our proposed learned designs (right). The respective LED pattern designs are shown below the phase reconstructions. Our learned LED source patterns enable improved reconstruction quality without sacrificing temporal resolution.}
    \label{fig:teaser}
  \end{figure}
}
\makeatother

\setcounter{figure}{-2}    
\maketitle

\section{Introduction}
\label{sec:intro}

\begin{figure}[t]
    \centering
    \includegraphics[width=8.89cm]{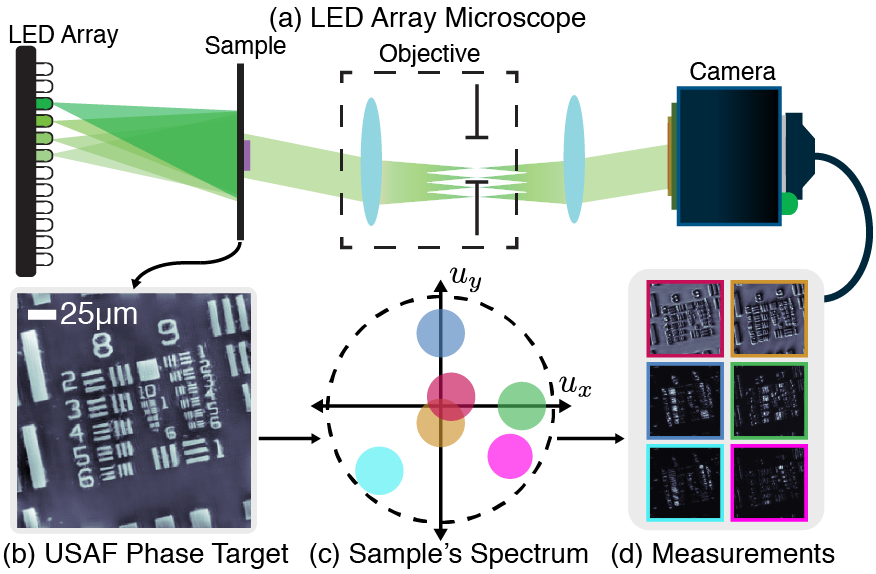}
    \caption{Fourier Ptychographic Microscopy (FPM) with an LED array source: (a) Schematic of experimental setup. (b) Test sample is a USAF phase target. (c) Each LED modulates a different part of the sample's Fourier space into the passband of the microscope. The colored circles correspond to different LED's Fourier space support, each of which has a small bandwidth (low resolution). The outer dashed circle represents the final super-resolution reconstruction's Fourier space support. (d) Intensity images are captured in real space by the camera sensor under different LED illuminations.}
    
    \label{fig:fpmSetup}
\end{figure}

\IEEEPARstart{F}{ourier} Ptychographic Microscopy (FPM)~\cite{Zheng:2013gq} is a computational imaging method that achieves both large field-of-view (FOV) and high resolution for both amplitude and quantitative phase imaging (QPI), enabling high-throughput imaging for applications in pathology~\cite{Zheng:2013gq} and live cell imaging~\cite{Tian:2015er}. It can be conveniently implemented using an inexpensive hardware modification to a conventional microscope simply by replacing the illumination source with a programmable LED array~\cite{Zheng:2013gq,Phillips:17} (Fig.~1). By capturing many low-resolution images, each using a different LED to encode a distinct part of the sample's Fourier space (Fig.~\ref{fig:fpmSetup}c), the high-resolution complex transmittance function of the sample can be recovered via non-linear phase retrieval optimization.

FPM increases the space-bandwidth product of the microscope, but comes at the cost of significantly reduced temporal resolution. In the originally proposed FPM system, a single measurement is acquired per LED~\cite{Zheng:2013gq} (single-LED design). To produce a high-quality reconstruction, measurements must redundantly encode information such that neighboring LEDs result in at least $60\%$ overlapping coverage of the sample's Fourier space~\cite{Sun:16,BUNK2008481}. This results in a poor data input:output ratio - approximately $10\times$ more pixels are acquired than the number of pixels reconstructed~\cite{Tian:2015er}. In practice, dozens of measurements are used and the slow speed of capture prevents imaging of live cell dynamics. Several works have improved upon the temporal resolution of the single-LED design, by turning on multiple LEDs simultaneously~\cite{Tian:2014wv,Tian:2015er}, spatially multiplexing the measurements across the sensor~\cite{Sidorenko:2016in}, color-multiplexing~\cite{dong2014spectral}, only acquiring the most important measurements~\cite{bian2014content}, or motion correction~\cite{kellman2018motion,bian2016motion}.

{\it Multiplexed} FPM~\cite{Tian:2014wv,Tian:2015er} improves upon the measurement requirement for single-LED design to achieve a data input:output ratio of approximately $1\times$ without sacrificing FOV or resolution. This is accomplished by acquiring fewer measurements, each taken under multiple-LED illumination, thereby encoding multiple parts of the sample's Fourier space into each measurement. However, as fewer measurements are acquired and more of the sample's Fourier space is encoded per measurement, the reconstructions become blurry. Practically, the reconstruction's performance is governed by the system's experimental design: which LEDs are turned on in each measurement and how many measurements are acquired.


Recent work has shown that supervised learning can be used to find the experimental design that optimizes the performance of a computational imaging system. Several model-free methods~\cite{Robey:2018vq,8389203} consider learning design parameters in addition to a black-box Neural Network (NN) that learns an image reconstruction. These methods require a huge number of training examples to properly learn millions of parameters that model the reconstruction process, and they often do not transfer well to the experimental setting. In comparison, model-based methods~\cite{Kellman:2018tz,Sitzmann:2018bd} are able to efficiently learn the experimental design with very few training examples and have been shown to learn designs that do transfer well to the experimental setting. This is achieved by {\it unrolling} the image reconstruction process~\cite{Gregor:2010,sun2016deep,Diamond:2017wa,7944641} into a Physics-based Neural Network (PbNN)~\cite{Kellman:2018tz} and then learning the experimental design parameters to maximize system performance.


In this work, we use the physics-based learned design framework~\cite{Kellman:2018tz} to find LED pattern designs that maximize the reconstruction performance of the FPM system. Using our learned designs, we demonstrate the ability to reduce the data input:output ratio below 1 without compromising reconstruction performance, thereby improving temporal resolution of the system, without sacrificing FOV or resolution. Next, by incorporating a context-specific loss function, we are able to tailor the learned LED patterns to different applications ({\it e.g.} amplitude contrast imaging vs. QPI). In experiment, we demonstrate that the LED patterns learned in simulation work well for both amplitude contrast imaging and QPI applications. Finally, we discuss interpretations of our learned designs and confirm that they match our intuition, as well as practical implementation details required to build PbNNs for large-scale systems.

\begin{figure*}[tb]
    \centering
    \includegraphics[width=18.19cm]{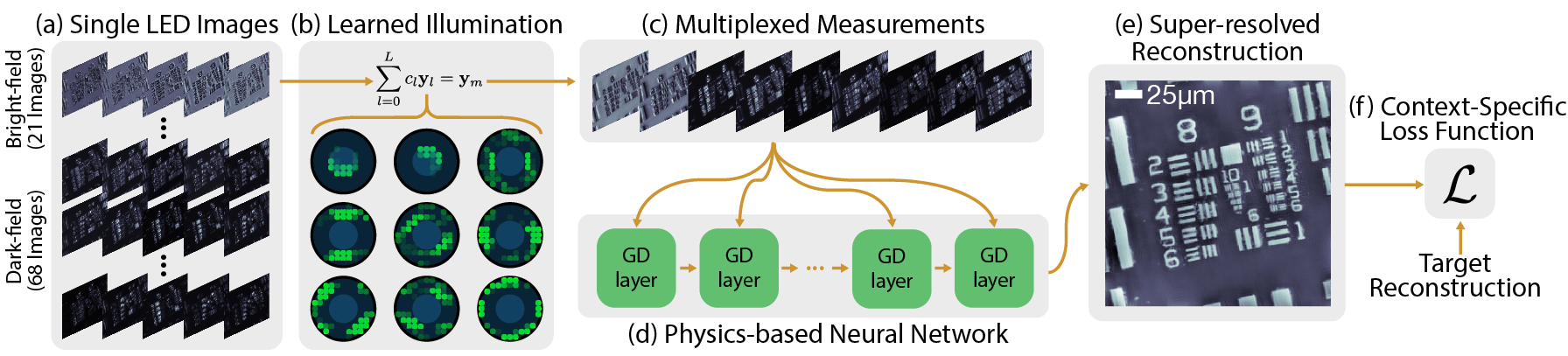}
    \caption{Physics-based Learned Design: (a) We start with an input dataset of single-LED images. (b) For each measurement's LED pattern, we learn the scalar weights, $c_l$, that correspond to the brightness of each LED. (c) Each multiplexed measurement is then fed into (d) our Physics-based Neural Network (PbNN), where each layer represents a single gradient descent (GD) step of the image reconstruction. (e) The output super-resolved reconstruction is then compared to the target reconstruction via (f) a context-specific loss function, in order to optimize the brightness of each LED.}
    \label{fig:PbNN}
\end{figure*}

\section{Fourier Ptychographic Microscopy}
\label{sec:fpm}
As mentioned in Section~\ref{sec:intro}, FPM achieves super-resolution, in that it resolves features beyond the diffraction limit of the microscope's objective. To do so, a set of low-resolution measurements recorded under varying illumination configurations are combined using a computational reconstruction. In this section, we revisit the principles of FPM and mathematically describe how a complex-valued reconstruction is obtained.

\subsection{Forward Model}
\label{ssec:forward}
Given the optical setup in Fig.~\ref{fig:fpmSetup} --- a conventional microscope with a coded LED source --- we model the optically-thin sample by its 2D transmittance function, $x(\rcoord) = e^{\cj \phi(\rcoord) - \mu(\rcoord)}$, where $\phi(\rcoord)$ and $\mu(\rcoord)$ represent the phase and absorption distributions of the sample, respectively, and $\rcoord \in \mathbb{R}^2$ is the spatial coordinate vector.

Each individual LED's illumination at the sample plane is modeled as a monochromatic plane wave (with wavelength $\lambda$) propagating at an angle characterized by the LED's physical position with respect to the azimuth of the microscope~\cite{Zheng:2013gq}. When the $l$th LED is ``on'', the complex-field after the sample can thus be written as $x(\rcoord) \exp \left( 2 \pi \cj \langle \boldsymbol{\xi}_l, \rcoord \rangle \right)$, where $\boldsymbol{\xi}_l \in \mathbb{R}^2$ is the spatial frequency vector corresponding to the angle of illumination. This can be interpreted as a translation in the Fourier domain (pupil plane),
    
    \begin{equation}
        \Fo \lbr x(\rcoord) \rbr (\mathbf{u} - \boldsymbol{\xi}_l) = \Fo \lbr x(\rcoord) \exp \left( 2 \pi \cj \langle \boldsymbol{\xi}_l, \rcoord \rangle \right) \rbr,
    \end{equation}
    
\noindent where $\mathbf{u}$ is the 2D spatial frequency vector.
    
At the pupil plane of the microscope, the sample's shifted Fourier space is low-pass filtered due to the finite-aperture objective lens, then Fourier-transformed again by the 2nd (tube) lens of the microscope. At the camera, an intensity-only (\textit{i.e.} phaseless) real-space image is captured. Mathematically, the process relating the sample's transmission function to the measured low-resolution image $y_l(\rcoord)$ is expressed as
    
\begin{equation}
    y_l(\rcoord) = \Bigl \lvert \Fo^{-1} \lbr  P(\mathbf{u}) \,
                    \Fo \lbr x (\rcoord) \rbr (\mathbf{u} - \xibf_{l}) \rbr  \Bigr\lvert^2 \text{.} 
\end{equation}

\noindent Here, $\Fo$ and $\Fo^{-1}$ denote Fourier transform and its inverse, respectively, $P$ is the \textit{pupil function} of the microscope objective, which has a hard frequency cutoff at the diffraction limit set by its numerical aperture ($\|\mathbf{u}\|_2 < \frac{\text{NA}_{\text{obj}}}{\lambda}$)~\cite{Goodman.1996}.
    
In {\it multiplexed} FPM (where $L$ LEDs illuminate the sample simultaneously), the measured intensity is the weighted sum of the individual LED measurements, since LEDs are mutually incoherent with each other~\cite{Tian:2014wv,Tian:2015er}:

    \begin{equation}
        y_{\mathrm{m}}(\rcoord) = \sum_{l=1}^L c_l y_l(\rcoord),
        \label{eq:multiplexedMeasurement}
    \end{equation}

\noindent where the non-negative scalar $c_l$ represents the relative brightness of the $l$th LED. The implication of Eq.~\eqref{eq:multiplexedMeasurement} is that multiplexing mixes images, each of which non-linearly encodes information from a distinct region in the sample's Fourier space (see Fig.~\ref{fig:fpmSetup}), thereby reducing the number of required measurements and improving the temporal resolution over the single-LED design. However, there is a limit: as more LEDs are used per measurement, it becomes harder to retrieve the high-resolution complex image. It has been shown that the data input:output ratio can be reduced to approximately $1$, without a significant reduction in reconstruction quality~\cite{Tian:2015er,Tian:2014wv}.
    
Depending on an LED's illumination angle, its resulting intensity image will either be a bright-field or dark-field image~\cite{liu2014real}. Bright-field images (see Fig.~\ref{fig:PbNN}a) are produced by central LEDs (inner circle of LED patterns in Fig.~\ref{fig:PbNN}b) whose spatial-frequency vectors are within the acceptance angle of the objective (where $\|\boldsymbol{\xi}\|_2 < \frac{\text{NA}_{\text{obj}}}{\lambda}$). These measurements encode a sample's low spatial-frequency information and are generally very bright, with high signal-to-noise ratio (SNR). Dark-field images are produced by LEDs in the outer shell of patterns in Fig.~\ref{fig:PbNN}b (where $\|\boldsymbol{\xi}\|_2 > \frac{\text{NA}_{\text{obj}}}{\lambda}$). These measurements encode a sample's high spatial-frequency information and are generally very dim, with low SNR. Because bright-field and dark-field images have orders-of-magnitude different brightness, they will be affected by Poisson noise differently~\cite{Yeh:2015ey}, and so we always design their LED patterns separately.

\subsection{Inverse Problem}
\label{ssec:inverse}

Given a forward model described above, the sample's complex-field is reconstructed by solving a non-linear inverse problem. Following convention, we discretize the 2D transmission function and multiplexed measurements as $\x \in \mathbb{C}^q$ and $\y_\mathrm{m} \in \mathbb{R}^p$, respectively. Note that $p<q$ since the reconstructed transmission function has a higher space-bandwidth product than that of the measurements~\cite{Zheng:2013gq}.

Reconstruction of the super-resolved complex transmittance function is then cast as an optimization problem:
    
    \begin{equation}
        \x^{\star} = \arg \underset{\x}{\min} \sum_{k=1}^{K} \Bigl\|\y_{\mathrm{m}_k} - \sum_{l \in \Omega_k} c_l | \A_l\x|^2 \Bigr\|_2^2 \hspace{-.23em} \text{,}
        \label{eq:recon}
    \end{equation}
    
\noindent where $\Omega_k$ is the index set for LEDs that are ``on'' during the $k$th multiplexed measurement $\y_{\mathrm{m}_k}$. In Eq.~\eqref{eq:recon}, the discretized system model is given by

\begin{equation}
    \A_l =  \mathbf{F}^{\rm{H}} \mathbf{P}_l \mathbf{F}\text{,}
\end{equation}

\noindent where $\mathbf{P}_l \in \mathbb{C}^{p \times q}$ is the matrix representation of the pupil function shifted by $\xibf_l$, and $\mathbf{F}$ and $\mathbf{F}^{\rm H}$ represent the discrete (normalized) Fourier transform matrix and its inverse, respectively. In this work we solve the optimization problem via gradient descent for which the convergence of the iterates to a stationary point is established~\cite{Yeh:2015ey,bostan2018accelerated}.

\section{Physics-based Learned Design}
\label{sec:ddd}

Physics-based Learned Design \cite{Kellman:2018tz} formulates the experimental design as a supervised learning problem that optimizes the source patterns to maximize the performance of the system. The method takes advantage of the PbNN, a rethinking of iterative optimization procedures as NNs that incorporate known quantities such as the system model and reconstruction non-linearities. The inclusion of these quantities provide the robustness and generality associated with model-based reconstruction and reduces the number of learnable parameters, thereby significantly reducing the number of training examples required. In this section, we show how to utilize this framework for FPM, discuss tailoring the experimental design for a specific application, and analyze memory limitations.

\subsection{Physics-based Neural Network}
\label{ssec:network}

Our PbNN is defined by the gradient-based optimizer used to solve the inverse problem in Eq.~\ref{eq:recon} (Fig.~\ref{fig:PbNN}d). Rather than using a network made up of millions of learnable parameters ({\it e.g.} convolutional filters), which would require a large number of training examples, our network is constructed from well-defined functions ({\it i.e.} gradient operations and the system model). Each layer of our network corresponds to a single iteration of our gradient-based optimizer. The network,
    
\begin{equation}
    \x^{\star} = R_{\C}(\{\y_l\}_{l=1}^{L}),
\end{equation}
    
\noindent takes as input a set of single LED measurements $\{\y_l\}_l$ (Fig.~\ref{fig:PbNN}a), linearly combines them according to the scalar weights in $\C \in \mathbb{R}^{K \times L}$ (Fig.~\ref{fig:PbNN}b) (which represent the relative brightness for each of $L$ LEDs across $K$ measurements), and outputs the reconstructed super-resolved complex image, $\x^{\star}$ (Fig.~\ref{fig:PbNN}e). The only learnable parameters in the PbNN are the scalar weights that set the relative brightness of each LED, thereby enabling learning with just a few training examples. 
    
    

\begin{figure*}[t!]
    \centering
    \includegraphics[width=18.19cm]{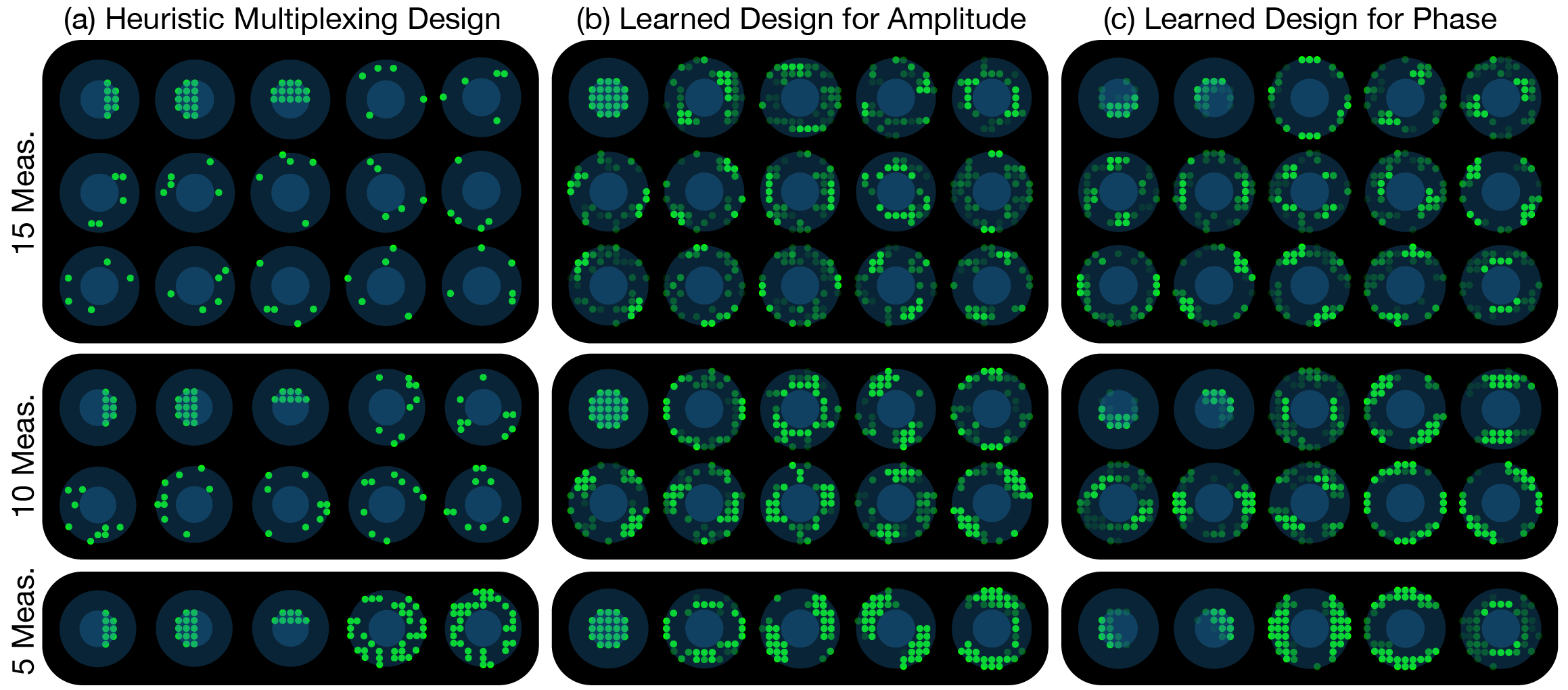}
    \caption{LED Designs: LED source patterns for (a) heuristic multiplexing designs with $15$, $10$ and $5$ measurements, (b) our learned designs for super-resolved amplitude imaging with $15$, $10$ and $5$ measurements, and (c) our learned designs for super-resolved quantitative phase imaging with $15$, $10$ and $5$ measurements. The inner blue circle denotes LEDs in the bright-field region, while the outer blue shell denotes the dark-field region. An LED's shade of green corresponds to that LED's brightness.}
    \label{fig:LED_designs}
\end{figure*}

\subsection{Context-Specific Learning Objective}
\label{ssec:context}

For specific applications, we can learn a custom experimental design by tailoring the loss function and training dataset. For example, with stained pathology slides, only amplitude contrast imaging is desired and phase is not used. In this case, our loss function penalizes only the amplitude error of the reconstruction. For live cell imaging applications, samples are nearly transparent and essentially pure phase objects. Hence, our loss function penalizes just the phase error of the reconstruction. For the general case, we can design for a spectrum of applications using this non-convex loss function,
    
    \begin{align}
        \mathcal{L}(\C) &= \sum_{n=1}^N \gamma \||\x_n^{\star}(\C)| - |\tilde\x_n|\|_2^2 \nonumber \\
        &+ (1 - \gamma) \|\angle \x_n^{\star}(\C) - \angle \tilde\x_n\|_2^2\text{,}
    \end{align}
    
\noindent where $\tilde\x_n$ is the ground truth complex transmittance function for the $n$th training example, $N$ is the total number of training examples, $\gamma \in [0,1]$ weights the loss between the phase ($\gamma=0$) and amplitude ($\gamma=1$)  loss functions, and $|\cdot|$ and $\angle$ return the amplitude and phase of the complex image, respectively.
    
In addition to tailoring the loss function, we also tailor our training examples to be predominately amplitude or phase samples, simulated from stock images of cells. For pathology applications, the dominant structural contrast is simulated in the absorption component of the sample's transmittance function. For QPI, the dominant structural contrast is in the phase component. Thanks to the efficient parameterization of the network by only the design parameters, we do not require a large amount of training data ($90$ image patches).
 
The loss function is then minimized via stochastic gradient descent (SGD) subject to several practical constraints,
    
\begin{align}
    \C^{\star} = &\arg\underset{\C}{\min} \ \mathcal{L}(\C) \quad \text{s.t.}\\
    & \mathbf{b}_k \odot \mathbf{c}_k = \mathbf{0} \,\,  & \text{(geometric)} \label{eq:phasecon} \\
    & c_{kl} \geq 0 \ \forall l \in \{1 \hdots L\}, \,\,   & \text{(non-negativity)} \label{eq:positive} \\
    & \|\mathbf{c}_k\|_1 = 1 & \,\,  \text{(scaling)} \label{eq:scaling} \\
    & \forall k \in \{1 \hdots K\}\text{,} \nonumber
\end{align}
    
\noindent where $\odot$ is element-wise multiplication. The geometric constraint (Eq.~\ref{eq:phasecon}) allows us to insert prior knowledge about the design by constraining certain LEDs to be ``on'' or ``off'' in each measurement. This is enforced using a mask, $\mathbf{b}_k$, for the $k$th measurement. In practice, we use the geometric constraint to separately design source patterns for the bright-field and dark-field regions. The non-negativity constraint (Eq.~\ref{eq:positive}) on the LED brightnesses enables our designs to be feasibly implemented in hardware. Finally, the scaling constraint (Eq.~\ref{eq:scaling}) removes degenerate solutions by eliminating arbitrary scalings of the same solution. After training, the overall brightness of the LED patterns can be scaled to utilize the full dynamic range of the camera or to a match desired image noise level.

\subsection{Computational Limitations}
\label{ssec:issues}

The feed-forward process of the PbNN has identical speed and memory complexity to that of its iterative reconstruction. However, when using graphical processing units (GPU) to accelerate the training process, the memory required for reverse-mode differentiation with backpropagation is limiting. The memory required is proportional to the product of the number of unrolled iterations, the number of measurements, and the size of the reconstructed image. To ensure convergence, we must use a sufficient number of unrolled iterations, which limits us to tuning only the latter two factors. To stay within our GPU's memory limit ($12$GB on an NVIDIA TITAN X Pascal GPU), we learn the design for reconstructing small patches ($35\text{px} \times 35\text{px}$) using up to $15$ measurements. Thanks to the Fourier relationship between FOV and sampling in Fourier space, our designs generalize well to larger FOVs.

\section{Results}
\label{sec:results}


We validate our learned designs for FPM with both simulated and experimental results. To learn the designs, we create a PbNN by unrolling $100$ iterations of gradient descent (as described in Sec.~\ref{fig:PbNN}) with the step size of $0.5$. Training is conducted in Pytorch using auto-differentiation to compute the gradients with respect to the learnable parameters. Data examples ($100$ patches each of $35\text{px} \times 35\text{px}$) are generated in simulation for two context-specific applications (amplitude contrast imaging and QPI). Each set of examples is shuffled and split into groups of $90$ and $10$ for training and testing, respectively. The training process is initialized using LED brightnesses drawn from a uniform random distribution.

Simulations are set up to match our experimental system parameters ($8\times$ objective, $\text{NA}_{\text{obj}} = 0.2$, with camera pixel size $\text{ps}_\text{camera} = 6.5\mu m$ and LEDs of wavelength $\lambda = 0.514\mu m$). We consider $89$ total LEDs on a Cartesian grid separated by a distance of $4mm$ in $x$ and $y$ with $21$ in the bright-field region and $68$ in the dark-field region, up to $\text{NA}_{\text{illum.}} = 0.42$. This allows us to reconstruct features up to $\text{NA}_{\text{recon.}} = \text{NA}_{\text{obj}} + \text{NA}_{\text{illum.}} = 0.62$. To mimic experimental noise, each measurement is simulated with a fixed exposure time such that the shot noise for bright-field measurements has a mean rate of $10$,$000$.


\begin{figure*}[tb]
    \centering
    \includegraphics[width=18.19cm]{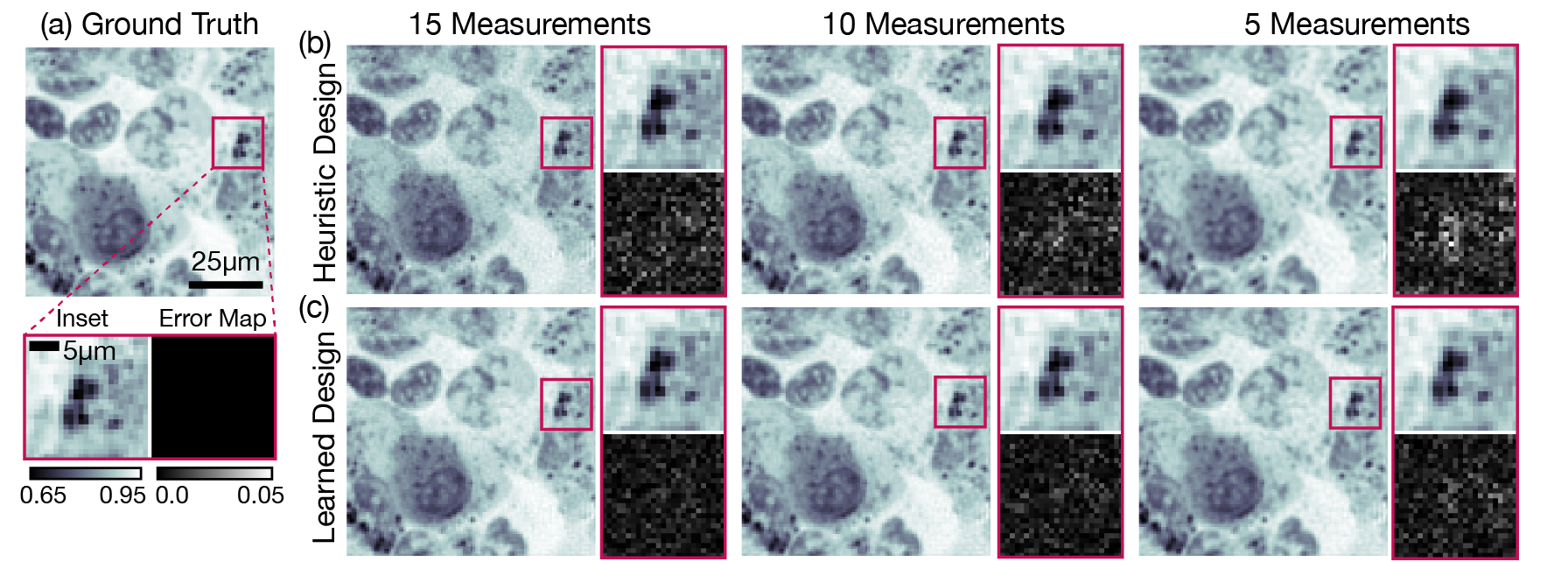}
    \caption{Simulations of super-resolution amplitude imaging: (a) Ground truth amplitude. (b) Amplitude reconstructions from simulated measurements using heuristic multiplexing designs and (c) our proposed learned designs, with $15$, $10$ and $5$ measurements. Insets highlight detailed features and their error maps.}
    \label{fig:amp_sim_sr}
\end{figure*}

\begin{figure*}[tb]
    \centering
    \includegraphics[width=18.19cm]{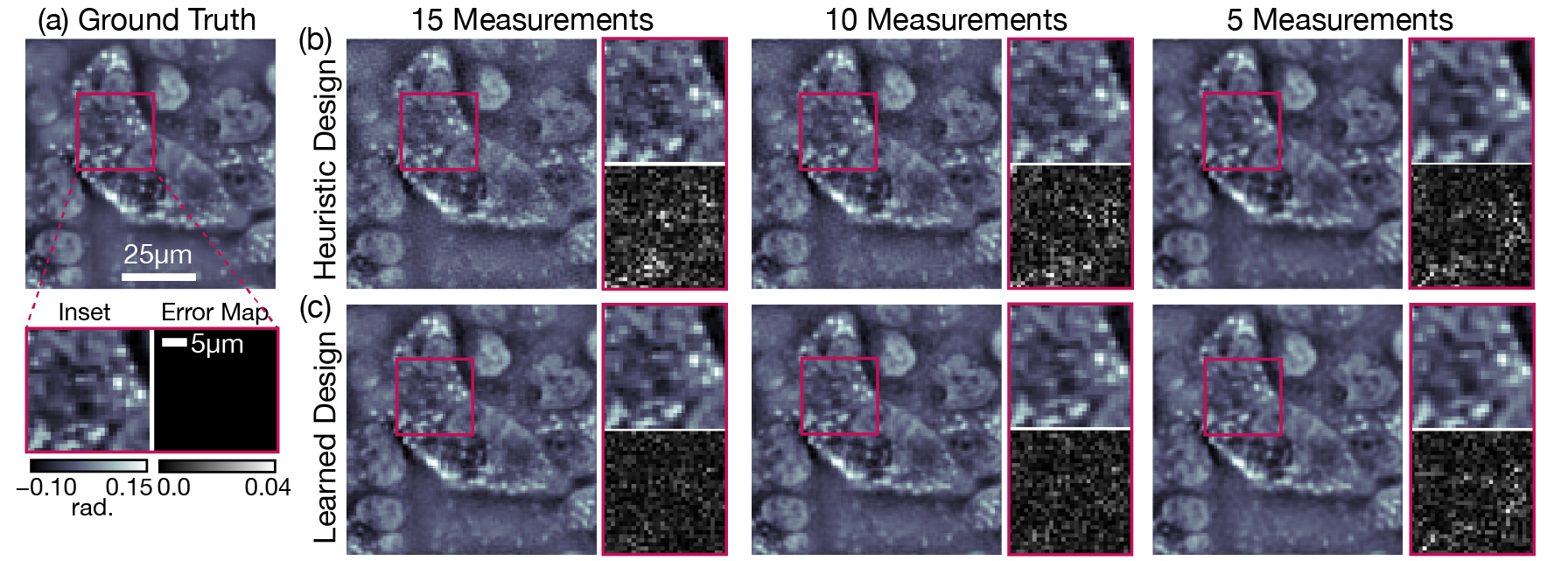}
    \caption{Simulations of super-resolution quantitative phase imaging: (a) Ground truth phase. (b) Phase reconstructions from simulations using heuristic multiplexing designs and (c) our proposed learned designs, with $15$, $10$ and $5$ measurements. Insets highlight detailed features and their error maps.}
    \label{fig:phase_sim_sr}
\end{figure*}

\subsection{Simulation Results}
\label{ssec:sim}

In Fig.~\ref{fig:LED_designs}, we display LED patterns for heuristic multiplexing designs for $15$, $10$, and $5$ measurements~\cite{Tian:2015er,Tian:2014wv}, as compared to our proposed learned designs for $15$, $10$, and $5$ measurements for both amplitude contrast imaging ($\gamma = 1$) and QPI ($\gamma=0$) applications. The heuristic multiplexing design for $K$ measurements consists of $3$ half-circle bright-field measurements and $K-3$ dark-field measurements. The LEDs in the dark-field measurements are randomly distributed such that each measurement has an equal number of LEDs ``on'' with equal brightness and each LED is ``on'' exactly once.

For the learned designs, our geometric constraint (Eq.~\ref{eq:phasecon}) enforces separate LED patterns for bright-field and dark-field LEDs. Specifically, we learn a single bright-field measurement for amplitude contrast imaging and two bright-field measurements for QPI. Generally, the amplitude contrast learned designs are symmetric in the bright-field, while the QPI learned designs are asymmetric in the bright-field (Fig.~\ref{fig:LED_designs}b,c). This makes sense because phase is anti-symmetrically encoded in Fourier space, whereas amplitude is symmetrically encoded. Unlike in heuristic multiplexing designs, the learned dark-field patterns are not random, but systematic. First, the learned designs are predominantly symmetric and LEDs are turned on in clusters, thereby encoding similar information of the sample's Fourier space in each measurement. Second, the LED clusters are fairly disjoint between measurements, thereby encoding different parts of the sample's Fourier space. Finally, as the number of measurements increases, the LED clusters decrease in size. These themes are further discussed in Section~\ref{sec:discussion}.

\begin{table}
    \centering
    \caption{PSNR for simulated amplitude reconstructions: average testing LF-PSNR and HF-PSNR (dB) for different numbers of measurements.}
    \begin{tabular}{ccc}
    \specialrule{.1em}{.05em}{.05em} 
            & Heuristic & Physics-based \\
            \# Meas. & Multiplexing &  Learned Design \\
            & (LF-PSNR/HF-PSNR) & (LF-PSNR/HF-PSNR)\\
            \hline
            15 & 47.84 / 16.15 & \bf{50.30 / 19.86} \\
            10 & 46.28 / 17.37 & \bf{49.39 / 19.83} \\
            5 & 43.83 / 18.40 & \bf{47.03 / 19.43} \\
            \specialrule{.1em}{.05em}{.05em}
        \end{tabular}
        
    \label{table:ampdBtable}
\end{table}
    

\begin{table}
    \centering
    \caption{PSNR for simulated phase reconstructions: average testing LF-PSNR and HF-PSNR (dB) for different numbers of measurements.}
    \begin{tabular}{ccc}
        \specialrule{.1em}{.05em}{.05em} 
        & Heuristic & Physics-based \\
        \# Meas. & Multiplexing &  Learned Design \\
        & (LF-PSNR/HF-PSNR) & (LF-PSNR/HF-PSNR)\\
        \hline
        15 & 31.01 / 14.97 & \bf{32.80 / 19.76} \\
        10 & 30.47 / 16.25 & \bf{31.85 / 19.88} \\
        5 & {\bf 29.94} / 19.42 & 28.93 / \bf{20.00} \\
        \specialrule{.1em}{.05em}{.05em} 
    \end{tabular}
    \label{table:phasedBtable}
\end{table}
    

\begin{figure*}[tb]
    \centering
    \includegraphics[width=18.19cm]{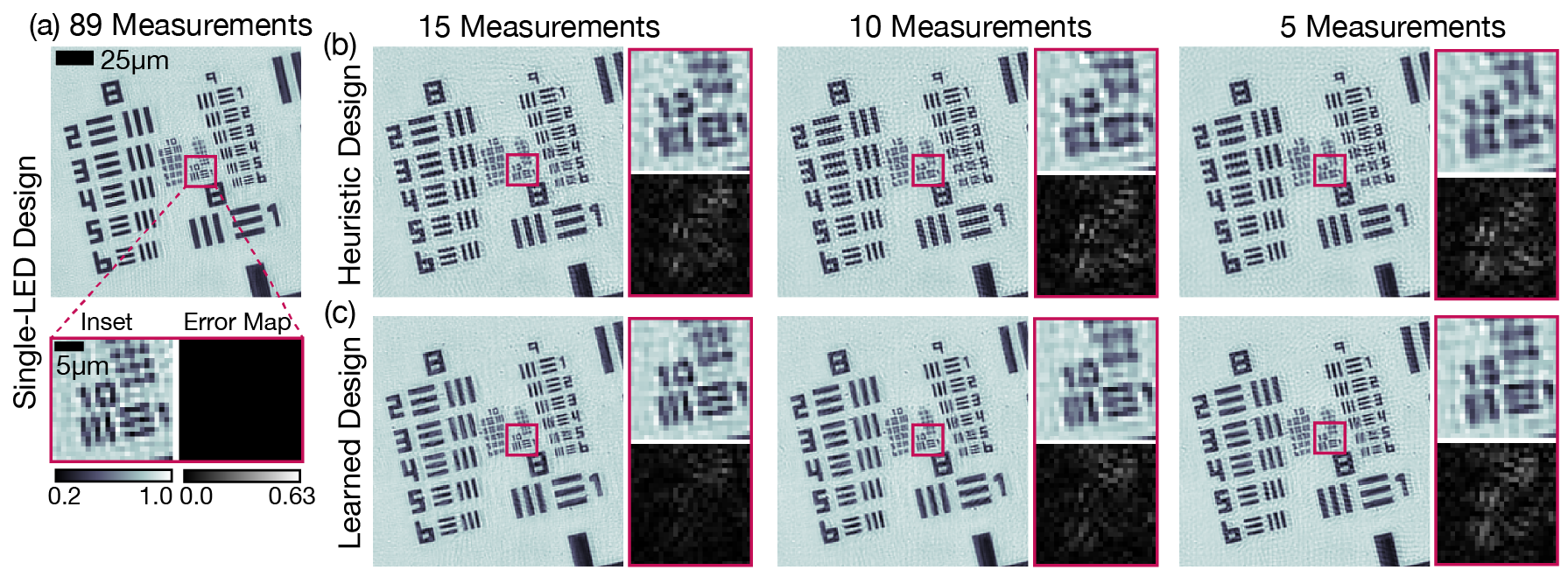}
    \caption{Experimental amplitude target results: (a) Single-LED amplitude reconstruction using $89$ measurements serves as ground truth. (b) Amplitude reconstructions using heuristic multiplexing designs and (c) our proposed learned designs, with 15, 10 and 5 measurements. Insets highlight detailed features and difference with the ground truth.}
    \label{fig:exp_amp_sr}
\end{figure*}

\begin{figure*}[tb]
    \centering
    \includegraphics[width=18.19cm]{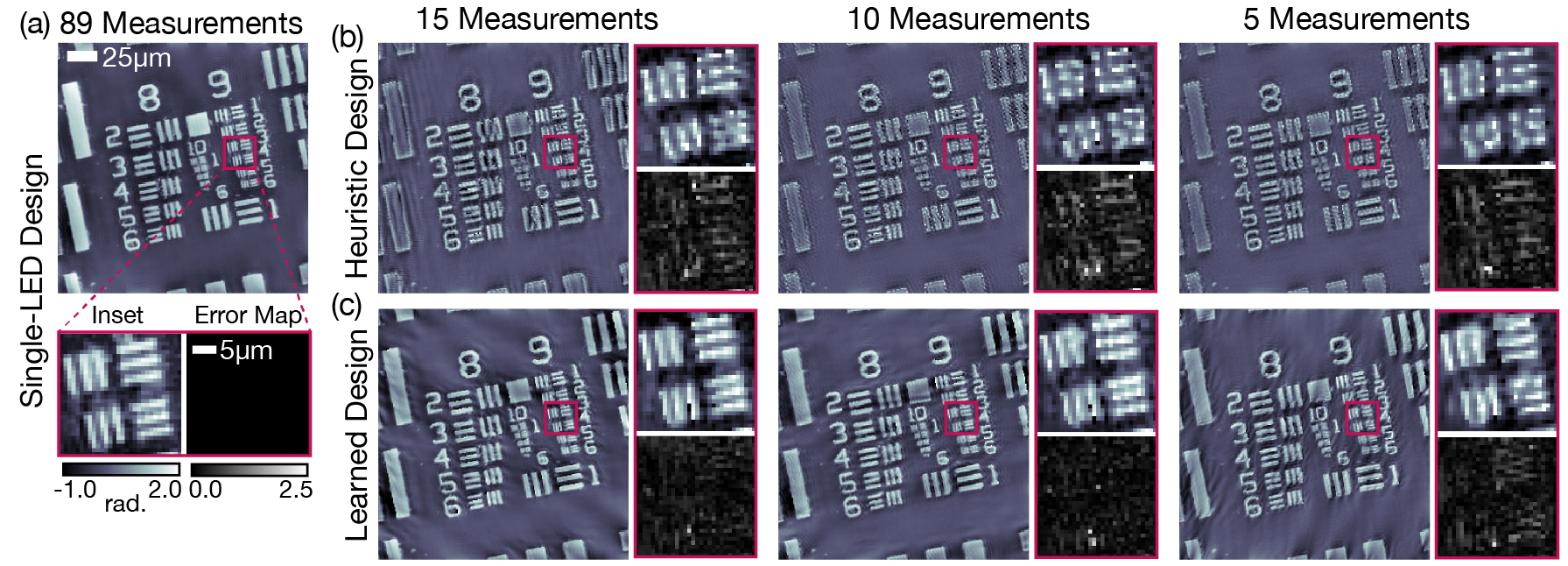}
    \caption{Experimental phase target results: (a) Single-LED phase reconstruction using $89$ measurements serves as ground truth. (b) Phase reconstructions using heuristic multiplexing designs and (c) our proposed learned designs, with 15, 10 and 5 measurements. Insets highlight detailed features and difference with the ground truth.}
    \label{fig:exp_phase_sr}
\end{figure*}


In Fig.~\ref{fig:amp_sim_sr} and Fig.~\ref{fig:phase_sim_sr}, we show super-resolved amplitude and QPI reconstructions from simulated measurements. We compare results for heuristic multiplexing designs~\cite{Tian:2015er,Tian:2014wv} and our learned designs to ground truth for $15$, $10$, and $5$ measurements (corresponding to data input:output ratios of $1$, $0.66$, and $0.33$, respectively). Insets highlight a small region and their differences with ground truth. Using heuristic multiplexing designs, as the number of measurements increases, finer detail features are resolved, but reconstructions become corrupted by artifacts. As the number of measurements decreases, reconstruction results become less noisy, but blurrier. In comparison, reconstructions using our learned designs remain sharp as the number of measurements decreases. 

To further quantify this trend, we report average testing peak signal-to-noise ratio (PSNR) for different regions of the reconstruction's Fourier space. Specifically, we calculate the PSNR for the low spatial-frequency region (LF-PSNR) up to the incoherent diffraction limit ($\text{NA} = 0.4$) and the PSNR for the high spatial-frequency region (HF-PSNR) ($\text{NA} = 0.4$ to $\text{NA} = 0.62$) to quantify high-frequency noise and the quality of super-resolved features. In Table \ref{table:ampdBtable} and Table \ref{table:phasedBtable} we report average testing LF-PSNR and HF-PSNR for super-resolved amplitude contrast imaging and super-resolved QPI applications, respectively. For heuristic multiplexing design, blurring causes the HF-PSNR for $5$ measurements to be relatively higher and noise causes the HF-PSNR for $15$ measurements to be relatively lower. In comparison, HF-PSNR for our learned designs are able to achieve approximately consistent performance as the number of measurements decreases. 

\subsection{Experimental Results}
\label{ssec:validation}

To validate our learned designs experimentally and show that training on simulated data is sufficient, we implemented our method on an LED array microscope. Our setup is a commercial Nikon TE300 microscope equipped with a custom LED array illumination system and a PCO.edge 5.5 monochrome camera ($2560\times2160$, $6.5\mu m$ pixel pitch, 16 bit). We image two samples: a USAF amplitude target and a USAF phase target (Benchmark Technologies). 

The experimental reconstructions of amplitude (Fig.~\ref{fig:exp_amp_sr}) and phase (Fig.~\ref{fig:exp_phase_sr}) targets compare our proposed learned designs to heuristic multiplexing designs~\cite{Tian:2015er,Tian:2014wv} and FPM single-LED design (89 measurements), which will serve as ``ground truth'' for validation. To make a fair comparison between different methods, we synthesize measurements for different designs by digitally combining a fixed set of single LED measurements. For amplitude contrast imaging (Fig.~\ref{fig:exp_amp_sr}), we show insets and their error maps, demonstrating resolution of features with a pitch of $0.97 \mu m$, even as the number of measurements is reduced. Heuristic multiplexing designs, on the other hand, lose quality and resolution as the number of measurements decreases. For QPI (Fig.~\ref{fig:exp_phase_sr}), we show insets and their error maps, demonstrating resolution of features with a pitch of $1.38 \mu m$, while reconstructions with heuristic multiplexing designs degrade with fewer measurements.


\section{Discussion}
\label{sec:discussion}

Physics-based learned design for FPM helps beat the trade-off between temporal resolution and reconstruction quality. Heuristic multiplexing designs have significantly improved temporal resolution, as compared to single-LED design, but reconstruction quality degrades as fewer measurements are acquired (data input:output ratio less than 1). Our learned designs produce systematic LED patterns that more efficiently encode the sample's Fourier space, thereby allowing for high-quality reconstructions using fewer measurements.

\begin{figure}
    \centering
\includegraphics[width=8.89cm]{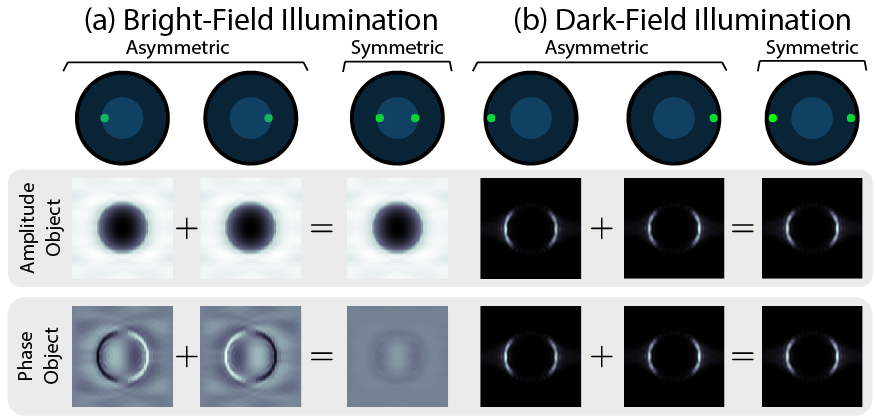}
\caption{Bright-Field vs. dark-Field LED designs: (a) Asymmetric and symmetric \textit{bright-field} LED patterns for amplitude and phase samples. (b) Asymmetric and symmetric \textit{dark-field} LED patterns for amplitude and phase samples.}
\label{fig:interp}
\end{figure}

For LED patterns in the bright-field region, our learned designs (Fig.~\ref{fig:LED_designs}) can be explained intuitively. Asymmetric illumination yields strong phase contrast and symmetric illumination yields strong amplitude contrast. In Fig.~\ref{fig:interp}a, we illustrate this by simulating measurements of a circular object with asymmetric and symmetric LED illumination. For an amplitude object, a pair of asymmetric LEDs produce identical contrast intensity measurements, such that when the sample is symmetrically illuminated the image contrasts add constructively. For a phase object, a pair of asymmetric LEDs produce opposite contrast intensity measurements, such that when the sample is illuminated the two image contrasts add destructively, reducing image contrast. Hence, phase samples will provide better contrast with asymmetric bright-field LED patterns, whereas amplitude samples favor symmetric patterns.

In contrast, our learned designs for dark-field LEDs always form symmetric clusters (Fig.~\ref{fig:LED_designs}). In Fig.~\ref{fig:interp}b, we simulate measurements of a circular object with asymmetric and symmetric dark-field LED patterns to understand why. For both amplitude and phase samples, pairs of asymmetric LEDs produce similar image contrasts such that when either sample is illuminated symmetrically the two image contrasts add constructively. We speculate that by turning on dark-field LEDs that produce similar contrast within the same measurement, it is less ambiguous where the encoded information is from in the sample's Fourier space and thus the reconstruction is able to retrieve the sample's high-resolution Fourier space more faithfully. Practically, symmetrical illumination increases the total illumination brightness, thereby allowing a reduction in the total acquisition time.





\section{Conclusion}
\label{sec:conc}

In this work, we have introduced a physics-based learned design framework to create interpretable context-specific LED source patterns for Fourier Ptychographic Microscopy, a highly non-linear computational imaging system. With these learned designs, we can achieve a more favorable trade off between temporal resolution and reconstruction quality, and tailor our source designs for context-specific applications such as amplitude imaging and quantitative phase imaging. Finally, we demonstrate that designs learned in simulation generalize well in the experimental setting.

\ifpeerreview
\else
\section*{Acknowledgment}

The authors thank Professor Michael Lustig for his thoughtful guidance and Dr. Emma Alexander for advanced readings.

\section*{Funding Information}

This work was supported by STROBE: A National Science Foundation Science \& Technology Center under Grant No. DMR 1548924 and by the Gordon and Betty Moore Foundation's Data-Driven Discovery Initiative through Grant GBMF4562 to Laura Waller (UC Berkeley). Laura Waller is a Chan Zuckerberg Biohub investigator. Michael R. Kellman is additionally supported by the National Science Foundation's Graduate Research Fellowship under Grant No. DGE 1106400. Emrah Bostan's research is supported by the Swiss National Science Foundation (SNSF) under grant P2ELP2 172278.

\fi

\bibliographystyle{IEEEtran}
\bibliography{ICCP2019}

\end{document}